\documentclass{amia}
\usepackage{graphicx}
\usepackage[labelfont=bf]{caption}
\usepackage[superscript,nomove]{cite}
\usepackage{color}
\usepackage{url}
\usepackage{subfigure}

\newcommand{\sks}{\mathcal{SK^S}}
\newcommand{\ske}{\mathcal{SK^E}}
\newcommand{\skaes}{\mathcal{SK^{AES}}}
\newcommand{\pks}{\mathcal{PK^S}}
\newcommand{\pke}{\mathcal{PK^E}}
\newcommand{\hash}{\mathrm{H}}

\begin{document}

\title{Secure and Trustable Electronic Medical Records Sharing using Blockchain}

\author{Alevtina Dubovitskaya, MS$^{1, 2}$, Zhigang Xu, MD$^3$, Samuel Ryu, MD$^3$, \\Michael Schumacher, PhD$^{1}$, Fusheng Wang, PhD$^{4}$}

\institutes{
    $^1$University of Applied Sciences Western Switzerland (HES-SO), Sierre, VS, Switzerland;\\ $^2$\'{E}cole polytechnique fé\'{e}d\'{e}rale de Lausanne  (EPFL), Lausanne, VD, Switzerland; \\
    $^3$Stony Brook Medicine, Stony Brook, NY, United States;\\
    $^4$Stony Brook University (SBU), Stony Brook, NY, United States\\
}

\maketitle

\noindent{\bf Abstract}
\textit{
Electronic medical records (EMRs) are critical, highly sensitive private information in healthcare, and need to be frequently shared among peers. Blockchain provides a shared, immutable and transparent history of all the transactions to build applications with trust, accountability and transparency. This provides a unique opportunity to develop a secure and trustable EMR data management and sharing system using blockchain. In this paper, we present our perspectives on blockchain based healthcare data management, in particular, for EMR data sharing between healthcare providers and for research studies. We propose a framework on managing and sharing EMR data for cancer patient care. In collaboration with Stony Brook University Hospital, we implemented our framework in a prototype that ensures privacy, security, availability, and fine-grained access control over EMR data. The proposed work can significantly reduce the turnaround time for EMR sharing, improve decision making for medical care, and reduce the overall cost.
}

\section*{Introduction}

Electronic medical records (EMRs) are critical but highly sensitive private information for diagnosis and treatment in healthcare, which need to be frequently distributed and shared among peers such as healthcare providers, insurance companies, pharmacies, researchers, patients families, among others.
This poses a major challenge on keeping a patient's medical history up-to-date. 
Storing and sharing data between multiple entities, maintaining access control through numerous consents 
only complicate the process of a patient's treatment.
A patient, suffering from a serious medical condition such as cancer, or HIV, has to maintain the long history of the treatment process and post-treatment rehabilitation and monitoring. Having access to a complete history may be crucial for his treatment: for instance, knowing the delivered radiation doses or laboratory results is necessary for continuing the treatment.

A patient may visit multiple medical institutions for a consultation, or may be transferred from one hospital to another.  According to the legislation \cite{directive199595,HIPAAPrivacy,FDA,45-cfr-46}, a patient is given a right over his health information and may set rules and limits on who can look at and receive his health information. If a patient needs to share his clinical data for the research purposes, or transfer them from one hospital to another, he may be required to sign a consent that specifies what type of data will be shared, the information about the recipient, and the period during which the data can be accessed by the recipient. 
This may be extremely difficult to coordinate, especially when a patient is moving to another city, region, or country and may not know in advance the caregiver or hospital where he will be receiving care later on.

 Even if the consent is provided, the process of transferring the data is time consuming, especially if sending them by post. Sending the patients' data via email over the Internet is not considered in most hospitals as this could impose security risk while the patient's healthcare records are in transit\cite{HIPAAPrivacy}. Ecosystems for health information exchange (HIE) such as CommonWell Health Alliance aim to ensure that the data form patient electronic health record are securely, efficiently and accurately shared nationwide in US. This implies that once providers receives an access to the patient's health information it is difficult to guarantee that a patient could receive independent opinions from different healthcare providers. Moreover, such ecosystems do not address the requirements in case of transferring data from one country to another. 
 
Data aggregation for research purposes also requires the consent unless the data are anonymized.  However, it has been shown that independent release of locally anonymized medical data corresponding to the same patient and originated from different sources (e.g., several healthcare institutions visited by the patient) could cause de-identification of the patient, and, therefore, violation of privacy\cite{Baig2011,DubovRep}.

Relying on centralized entity that would store and manage the patients' data and access control policies means having single point of failure and a bottleneck of the whole framework. It also requires either conducting all the operations (such as search, or anonymization) over encrypted data, or choosing a fully trusted entity that will have access to sensitive information about the patients. The former still requires management of large amounts of memory \cite{moore2014practical} and is not suitable for  hospital environment. The latter was proven to be very difficult to put in practice. An example of GoogleHealth wallet\cite{googleH} has shown that patients are concerned about their privacy and aware of the potential risk that their sensitive data might be misused.

Having access to a ledger - shared, immutable, and transparent history of all the actions that have happened to all the participants of the network (such as a patient modifying permissions, a doctor, accessing or uploading new data, or sharing them for research) overcome the issues presented above. By providing the tool to achieve consensus among distributed entities without relying on a single trusted party, blockchain technology 
will guarantee data security, control over sensitive data, and will facilitate healthcare data management for the patient and different actors in medical domain.
In the healthcare settings we can define a transaction as a process of creating, uploading or transferring EMR data that is performed within the connected peers. A set of transactions grouped at certain time is added to the ledger that records all the transaction and therefore represents the state of the network. The key benefits of applying the blockchain technology in healthcare are the following: verifiable and immutable transactions; tamper resistance, transparency, and integrity of distributed sensitive medical data. This is mainly achieved by employing consensus protocol and cryptographic primitives such as hashing and digital signatures.

The possibility of using blockchain for healthcare data management has recently raised a lot of attention in both industry and academia \cite{MedRec,gem,beninger2016pharmacovigilance,yue2016healthcare}
.  However, 
only one functioning prototype of a system that uses blockchain for medical data management has been proposed \cite{MedRec}. 
%
In our work, we focus on a practical implementation of a system that uses blockchain technology and can be integrated in clinical practice. We employ permissioned blockchain technology to maintain metadata and access control policy and a cloud service to store encrypted patients' data. Combining these technologies allows us to guarantee data security and privacy as well as availability with respect to the access control policy defined by the patient.

The contribution of the paper is twofold. First, we propose multiple scenarios of blockchain applications in healthcare and analyze existing technology implementations that could be used to put the scenarios in practice. 
Second, we present a framework for blockchain based data sharing for primary care of oncology patients under cancer treatment. We developed a prototype in collaboration with the Department of Radiation Oncology in a major US hospital. Therefore, the functionality of the prototype is expected to meet the requirements from medical practice perspective.

\section{Background on Blockchain}
Blockchain is a peer-to-peer distributed ledger technology that was initially used in the financial industry. 
\cite{HL_WP}. 
Based on how the identity of a user is defined within a network, one could distinguish between permissioned and permissionless blockchain systems. A permissionless system is one in which the identities of participants are either pseudonymous or even anonymous \cite{swanson2015consensus} and every user may append a new block to the ledger. 
In contrast, in case of a permissioned blockchain, the identity of a user is controlled by an identity provider. The identity provider is trusted to maintain access control within the network and the user's rights to participate in the consensus, or validate a new block. 
Next we introduce two most well-known implementations of the blockchain technology: Ethereum \cite{buterin2014ethereum} and Hyperledger \cite{HL_WP}.
\vspace{-0.2cm}
\subsection{Permissionless Blockchain Implementation}
Ethereum \cite{buterin2014ethereum} is an implementation of a permissionless programmable blockchain that allows any user to create and execute the code of arbitrary algorithmic complexity on the Ethereum platform: Ethereum Virtual Machine (EVM). 
``Accounts'' of two types could be created on EVM. Externally owned account (EOA) is an account controlled by a private key of a user. 
Contract account is the second type of accounts that can be seen as an autonomous agent that lives in the Ethereum execution environment and is controlled by its contract code: smart contract. Smart contract is used to encode arbitrary state transition functions, allowing users to create systems with different functionalities by transforming the logic of the system into the code.

Code execution in Ethereum must be paid. The transaction price limits the number of computational steps for the code execution in order to prevent infinite loops or other computational wastage. Users could participate in a consensus process to obtain the tokens to be paid for transaction execution. In Ethereum, the consensus is achieved by using a proof-of-work (PoW) mechanism. PoW is based on ``mining'': finding a nonce input to the algorithm so that the resulting hash of a new valid block satisfies certain requirements. These requirements set the difficulty threshold for the process of finding the nonce \cite{ethDocs}.

The difficulty threshold impacts the amount of energy to be spent to find such nonce. For example, the amount of energy used by Bitcoin mining is comparable to the Irish national energy consumption \cite{o2013bitcoin}. Existing PoW blockchains can achieve throughput of not more than 60 transactions per second without significantly affecting the blockchain's security \cite{gervaissecurity}. These two findings show that PoW can negatively impact the system scalability and overall throughput \cite{vukolic2015quest}.

Proof-of-Stake (PoS) \cite{PoS} and Proof-of-Burn (PoB) \cite{PoB}, or virtual mining mechanisms, have been recently proposed as alternatives to PoW. Instead of having participants mine by exchanging their wealth for computational resources (which are then exchanged for mining rewards), in virtual mining, participants could exchange their wealth directly for the ability to append a new block to the ledger \cite{bonneau2015sok}. For example, in PoS, selection of a participant that will create a new block is based on the amount of tokens owned by the participant, in PoB -- based on the amount of tokens ``burnt'' (sent to an unspendable address).
 However, providing a rigorous argument for or against the stability of virtual mining remains an open problem \cite{bonneau2015sok}.
\vspace{-0.2cm}
\subsection{Permissioned Blockchain Implementation}

In the case of a permissionned system, users do not have an incentive to cheat as their identity is revealed to the identity server. Moreover, participation in consensus management is restricted to a predefined set of users. This opens a possibility to use a state machine replication algorithm (such as PBFT \cite{castro2002practical}) as a consensus mechanism. Hyperledger  \cite{HL_WP} -- an implementation of a permissioned blockchaian -- is an open source blockchain initiative hosted by the Linux Foundation. Hyperledger has a modular architecture that allows plugging in different consensus mechanisms, including PBFT. Hyperledger services could be logically grouped in three categories: Membership services, Blockchain services, and Chaincode services \cite{cachin2016architecture}.

Membership services manage identity, privacy, and confidentiality on the network. A user is assigned a username and a password that will be used to issue the Enrollment certificate (ECert) to identify every registered user.
It is possible to use different Transaction certificates (TCert) associated with the same ECert for every transaction to ensure their unlinkability (a mapping between TCert and Ecert are only known to the membership service).
Blockchain services manage the distributed ledger through a peer-to-peer protocol built on HTTP/2.
In Hyperledger, smart contracts are implemented by the chaincode.
Chaincode services provide a secure way to execute smart contracts on validating nodes.

In Hyperledger, smart contracts are implemented by chaincode that consist of Logic and associated World State (State).
Logic of the chaincode is a set of rules that define how transactions will be executed and how State will change. State is a database that stores the information in a form of keys and values that are arbitrary byte arrays. The State also contains the block number to which it corresponds. Ledger manages blockchain by including an efficiently cryptographic hash of the State when appending a block. This allows efficient synchronization if a node was temporary off-line, minimizing the amount of stored data at the node \cite{HL_WP}.

\section{Potential Blockchain Applications in Healthcare}
 \label{Blockchain-in-health Scenarios}
 \begin{figure}[!t]
\centering
\includegraphics[scale=0.90, width=3.3in]{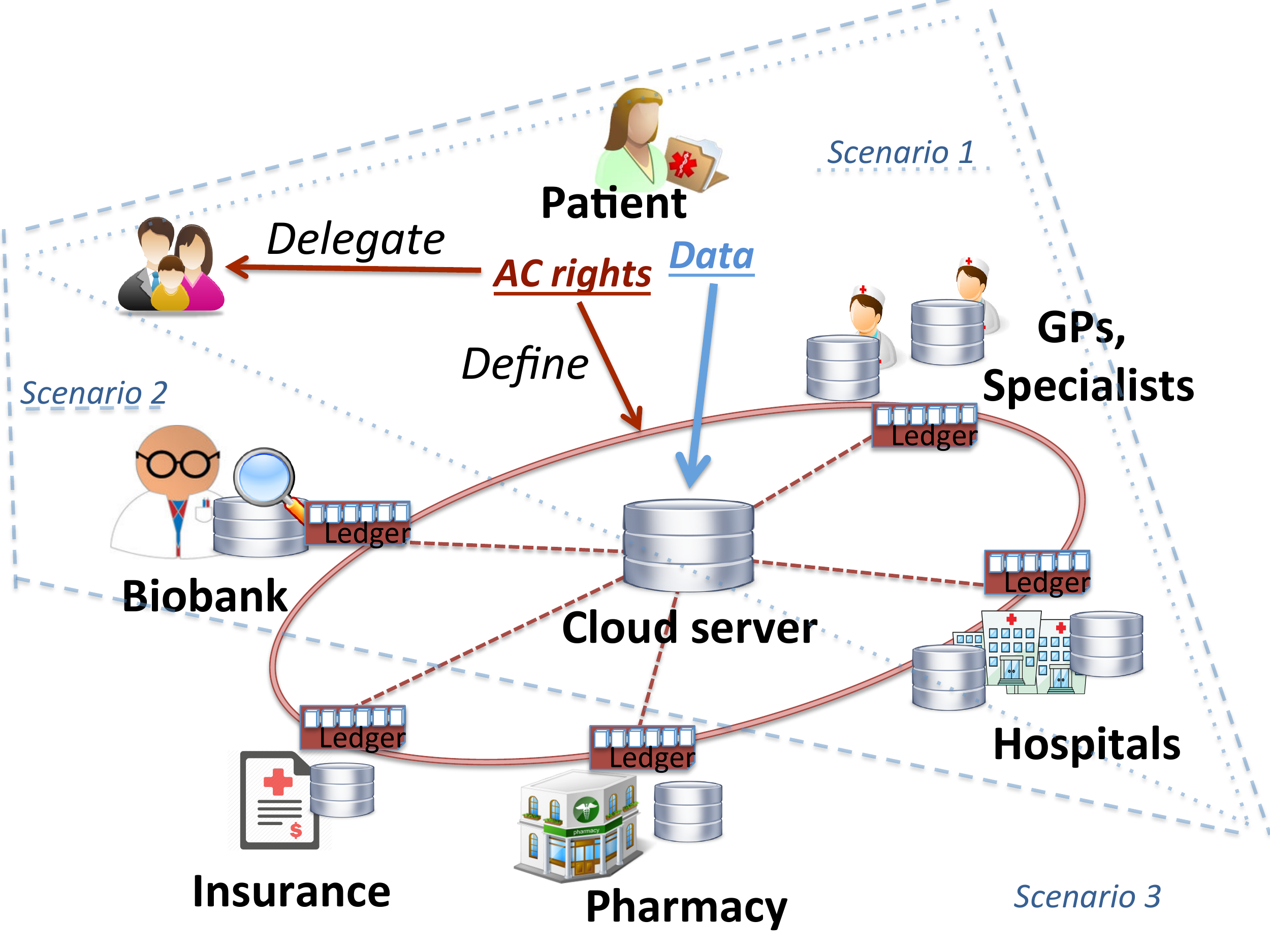}
\caption{Scenarios of using blockchain in different healthcare settings: Scenario 1: Primary patient care; Scenario 2: Data aggregation for the research purposes; Scenario 3: Connecting different healthcare players for better patient care.}
\label{scenarios}
\end{figure}
Blockchain provides a unique opportunity to support healthcare.  In this section, we propose three scenarios: primary patient care
, medical research
, and connected health
. Figure \ref{scenarios} shows a graphical representation of the combination of the aforementioned scenarios. 

\textbf{Scenario 1: Primary Patient Care.} Using blockchain technology for primary patient care can help to address the following problems of the current healthcare systems:
\begin{itemize}
\vspace{-0.2cm}
\item A patient often visits multiple disconnected hospitals. He has to keep the history of all his data and maintain the updates. This leads to the situation when required information may not be available.
\vspace{-0.2cm}
\item Due to the unavailability of the data, patient may have to repeat some tests for laboratory results. This is common when the results are stored in another hospital and can not be immediately accessed.
\vspace{-0.2cm}
\item The healthcare data are sensitive and their management is cumbersome. Yet, there is no privacy-preserving system in clinical practice that allows patients to maintain access control policy in an efficient manner.
\vspace{-0.2cm}
\item Sharing data between different healthcare providers may require major effort and could be time consuming.
\vspace{-0.2cm}
\end{itemize}
Next, we propose two approaches that can be implemented separately or combined to improve patient care.
\begin{itemize}
\vspace{-0.2cm}
\item \textit{Institution-based}: The network would be formed by the trusted peers: healthcare institutions or general practitioners (caregivers). The peers will run consensus protocol and maintain a distributed ledger. The patient (or his relatives) will be able to access and manage his data through an application at any node where his information is stored. If a peer is off-line, a patient could access the data through any other online node. The key management process and the access control policy will be encoded in a chaincode, thus, ensuring data security and patient's privacy.
\vspace{-0.2cm}
\item \textit{Case specific} (serious medical conditions, examination, elderly care): During a patient's stay in a hospital for treatment, rehabilitation, examination, or surgery, a case-specific ledger could be created. The network would connect doctors, nurses, and family to achieve efficiency and transparency of the treatment. This will help to eliminate human-made mistakes, to ensure consensus in case of a debate about certain stage of the treatment.
\end{itemize}

\textbf{Scenario 2: Data Aggregation for Research Purposes.} It is highly important to ensure that the sources of the data are trusted medical institutions and, therefore, the data are authentic. Using shared distributed ledger will provide tracebility and will guarantee patients' privacy as well as the transparency of the data aggregation process. Due to the current lack of appropriate mechanisms, patients are often unwilling to participate in data sharing. Using blockchain technology within a network of researchers, biobanks, and healthcare institutions will facilitate the process of collecting patients' data for research purposes. 
%

\textbf{Scenario 3: Connecting Different Healthcare Players for Better Patient Care. }
Connected health is a model for healthcare delivery that aims to maximize healthcare resources and provide opportunities for consumers to engage with caregiver and improve self-management of a health condition \cite{wiki}. Sharing the ledger (using the permission-based approach) among entities (such as insurance companies and pharmacies) will facilitate medication and cost management for a patient, especially in case of chronic disease management. Providing pharmacies with accurately updated data about prescriptions will improve the logistics. Access to a common ledger would allow the transparency in the whole process of the treatment, from monitoring if a patient follows correctly the prescribed treatment, to facilitating communication with an insurance company regarding the costs of the treatment and medications.

\textbf{Implementing the Scenarios.} In order to implement the three healthcare scenarios presented above, we must choose between a permissionless and a permissioned blockchain implementations. Below we present the facts that favor a permissioned system implementation. 
\begin{itemize}
\vspace{-0.2cm}
\item The anonymity of users and impossibility to verify the identity of account holders (as in case of permissionless blockchain) could cause impersonalization and data misuse.
\vspace{-0.2cm}
\item Patients' healthcare data are of high sensitive nature. Even monitoring communication between a patient and a specific clinician may reveal some sensitive data about the patient, therefore violating the privacy.
\vspace{-0.2cm}
\item Fast response of a system is required as any update of the information about a patient's treatment could be crucial for the patient.
\vspace{-0.2cm}
\item The need to pay for transaction execution, for example, updating permissions for a medical doctor to access a piece of healthcare information or sharing some data for research could limit the usability of the system.
\end{itemize}

\section{Application in Radiation Oncology: Sharing Clinical Data between Healthcare Providers}
In this section, we present a prototype design and implementation of a system to support electronic medical record sharing  for primary patient care (Scenario 1). More precisely, we focus on  patients that are receiving a cancer treatment via  ionizing radiation, which is usually performed in the Department of Radiation Oncology of a hospital.
First, we describe a specific use-case scenario and the benefits of the system.
Second, we present the architecture of the system and describe the data structure and functionality of the system.
Finally, we discuss how privacy, security, and scalability are ensured within the proposed framework.
\vspace{-0.2cm}
\subsection{Use Case Scenario}
Cancer is a serious medical condition that may require a long-lasting treatment and a life-time monitoring of a patient. Therefore, it is crucial for the patient to maintain his medical history and to be able to access or share his medical data during the treatment and post-treatment monitoring. Due to the mobility of a patient, the management of the data generated during every patient's visit can be cumbersome especially given the sensitive nature of healthcare data. How to guarantee that the patient's data are complete, stored securely, and can be accessed only according to the patient consent in a fast and convenient manner?

We tackle this problem by applying the blockchain technology to create a prototype of an oncology-specific clinical data sharing system. To present our solution, we take as an example an oncology information system, ARIA\cite{aria}, which is widely used to facilitate oncology-specific comprehensive information and images management. ARIA combines radiation, medical and surgical oncology information and can assist clinicians to manage different kinds of medical data, develop oncology-specific care plans, and monitor radiation dose of patients.
Different types of data stored in this system can be structured depending on the clinician's request and exported in PDF format. The documents that contain the data such as history and physical exams, laboratory results, and delivered radiation doses are of the high importance for the clinicians and are most commonly used during the treatment.

Currently, if any of these data have to be transferred from Hospital 1 to Hospital 2, the following procedure takes place.
First, the patient (or his official representative) has to sign a consent -- a document that specifies the data to be transferred and contains the information about the recipient of the data (Hospital 2).
Then, the information has to be printed and mailed to the recipient.
Consent management and data transfer in this case can become complicated and inconvenient: the patient may need to contact the caregiver and sign a consent in the hospital from which  he is not receiving care anymore. Data transfer can take time, and on receiving the hard copy of the patient data, a clinician will have to introduce them into 
the system again. Moreover, with this approach, it is very difficult for the patient to maintain any access control of his data and to have a complete view of the data.

By employing blockchain technology, our solution allows to facilitate the consent management
and speed up data transfer.
We developed a chaincode that allows a patient to easily impose fine-grained access control policy for his data and enables efficient data sharing among clinicians.
\vspace{-0.2cm}
\subsection{System Architecture}
\begin{figure}[!t]
\centering
{\includegraphics[width=4in]{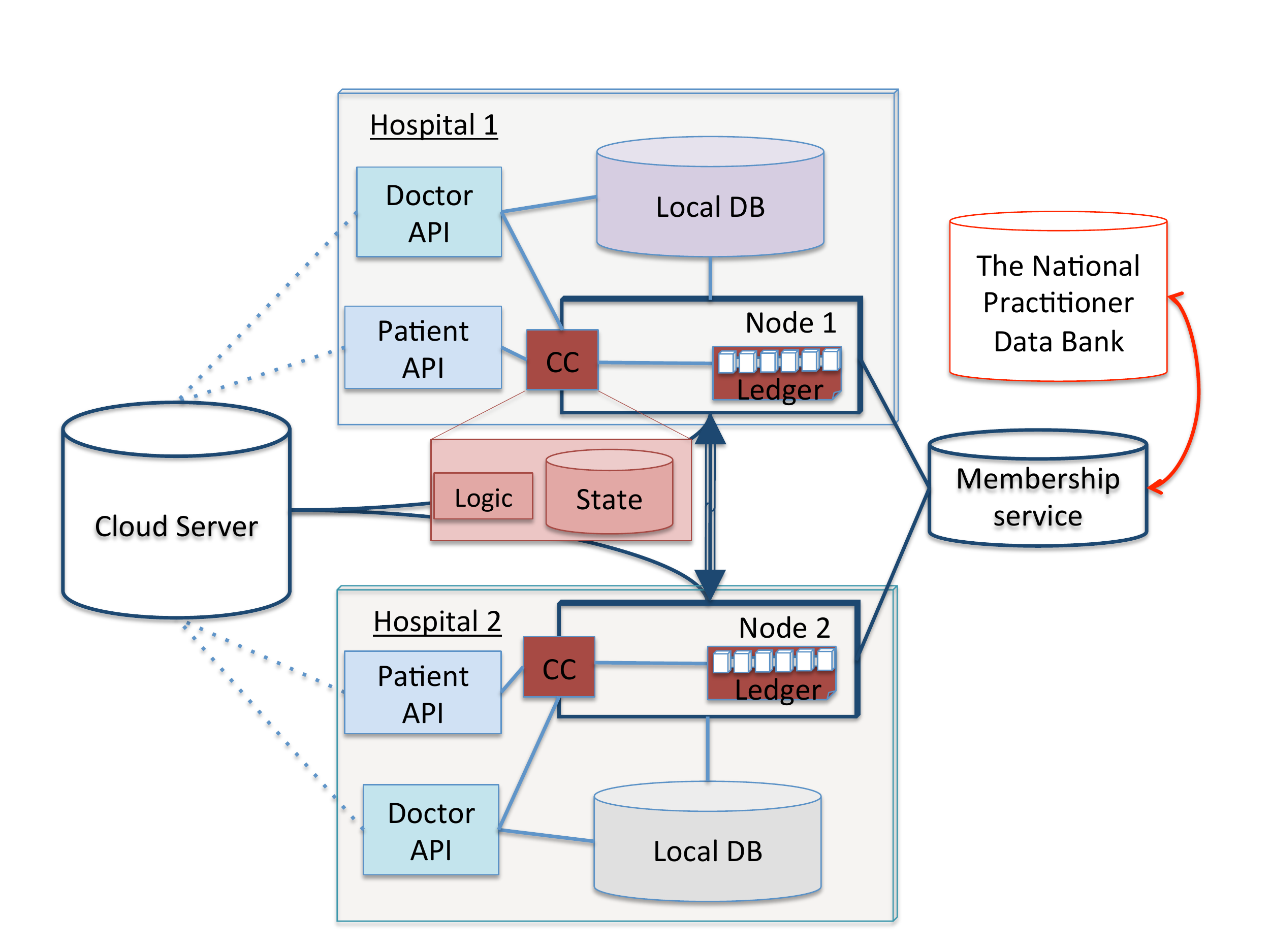}}
\caption{The System Architecture of Blockchain Based Data Management and Sharing for Radiation Oncology}
\label{architecture}
\end{figure}

Figure \ref{architecture} presents the architecture of our framework for the oncology-specific data management. The framework consists of the \textit{Membership service}, \textit{Databases} for storing healthcare data off-chain, \textit{Nodes} managing consensus process, and \textit{APIs} for different user's roles. Currently, we focus on Doctor and Patient, but the roles and their functionality could be extended depending on the scenario.

The main functionality of the \textit{Membership service}
is to register users with different roles (currently Doctor and Patient).
The roles define the functionality of the chaincode that is available to the user. During the registration of a user as a Doctor, it is important to ensure that it is not a potential malicious user, but a qualified medical doctor. To verify this, the National Practitioner Data Bank could be consulted by the membership service. The membership service is also hosting a certification authority involved in the generation of a key pair for signing ($\sks_{U}, \pks_{U}$) and encryption key pair ($\ske_{U}, \pke_{U}$) for every user ($U$).

Patient ($P$) also generates a symmetric encryption key ($\skaes_{P}$)
that will be used to encrypt/decrypt the data corresponding to the patient, $P$.
This key will also be used to generate pseudonyms so that only authorized users could verify whether the ledger stores any information about the patient.
When a patient ($P$) needs to share his data with a doctor ($D$), the patient could share this key, $\skaes_{P}$, using the encryption public key of the corresponding clinician ($\pke_{D}$). If the symmetric key $\skaes_{P}$ is compromised, the patient could generate a new one, run a proxy re-encryption algorithm \cite{Ateniese:2006} on the data stored in the cloud and then share a new key with the clinicians according to the desired access control policy.

The patient's data are stored off-chain in the following \textit{Databases}.
First, a local database management system in the hospital that stores the oncology-related data (for example, ARIA in our use case). Second, a cloud based platform (Varian Cloud) that stores patient's data organized based on the data category (for instance, according to the sensitivity level of the data, or their semantics), and encrypted with corresponding patient key, $\skaes_{P}$. 
A registered clinician could assess or upload the data in the cloud repository based on the access control policy defined by the patient and implemented in the chaincode Logic.

A custom chaincode is deployed on every \textit{Node} that acts as a Hyperledger validating peer. Nodes receive all transactions submitted by the users through a role-based \textit{APIs}. The Node, selected as a leader, organizes transactions in a block and initiates the PBFT consensus protocol. Transactions are executed by all nodes according to the implemented chaincode Logic.
The State stores the information about patients in a key-value pair format. A Key -- a Patient Id in the system -- is a pseudonym of the patient that is generated as a hash of the  concatenation of the symmetric key $\skaes_{P}$ and a Uniquely Identifiable Information of the patient ($\textsc{UII}_P$): $\hash \left( \skaes_{P} \parallel \textsc{UII}_P\right)$. Combination of SSN (if applicable), date of birth, given names, and a ZIP code of the patient could be used as $\textsc{UII}_P$.
A Value is a patient record stored as a byte array.
Next we describe the data structure in detail.

\subsection{Data Structure and System Functionality}

\begin{figure}[!t]
\centering
{\subfigure[The structure of the patient's data stored in the cloud.]{\includegraphics[trim=4cm 3cm 3cm 3cm, clip=true, width=2.1in]{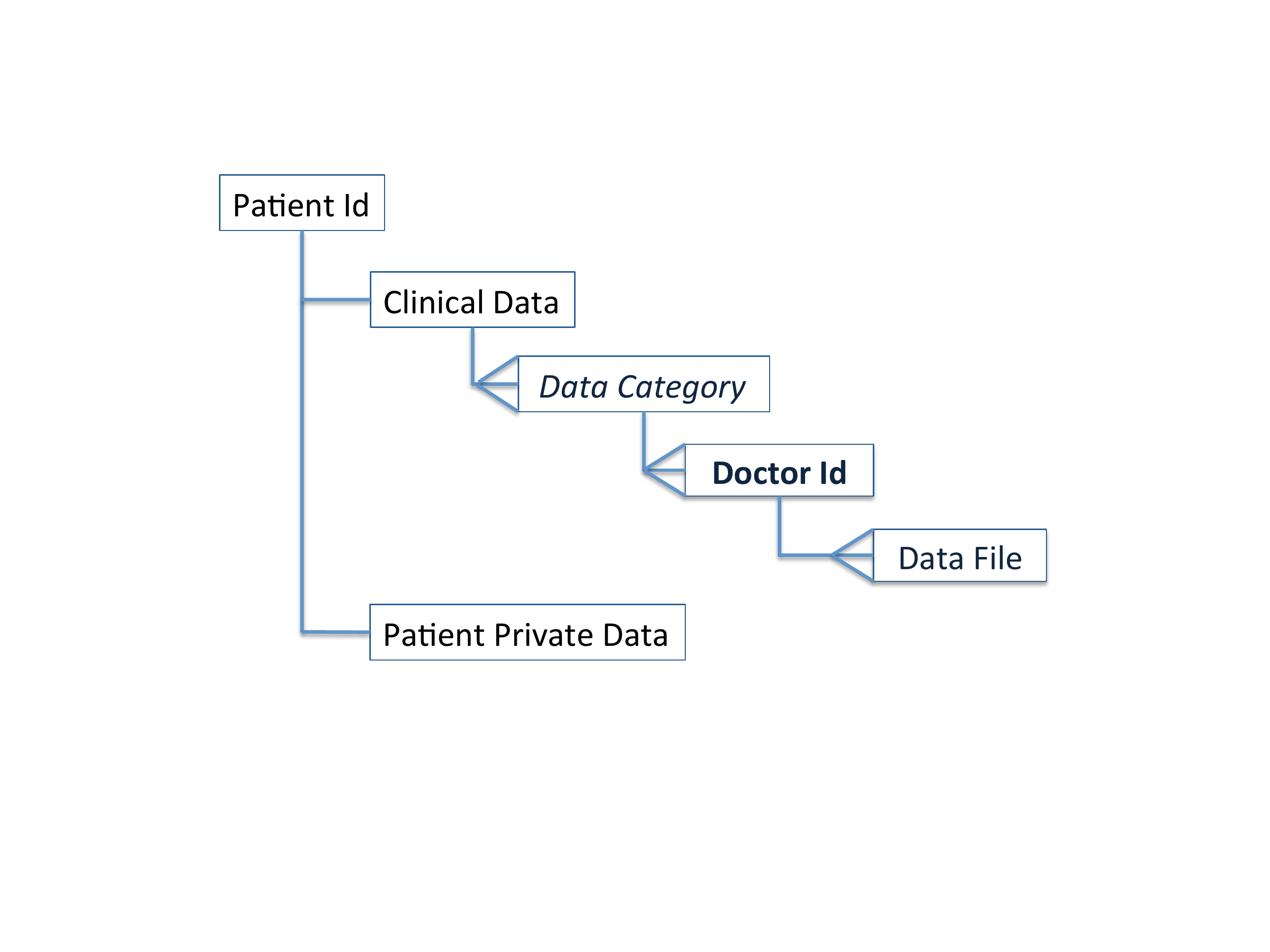}}
\subfigure[The structure of the patient's metadata stored on the chaincode.]{\includegraphics[trim=0cm 1cm 0.5cm 0cm, clip=true, width=2.6in]{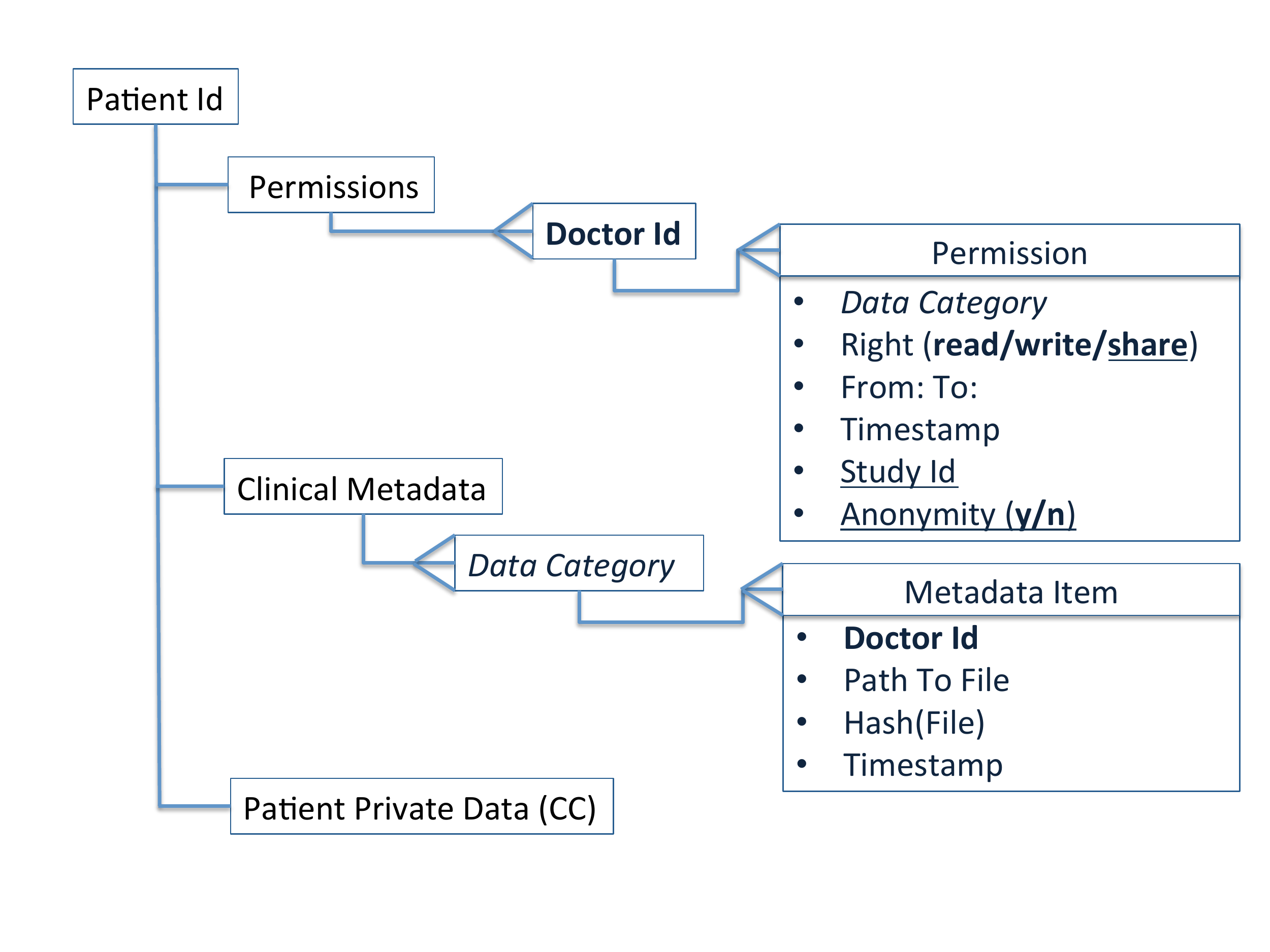}}}
\caption{The data structure of a patient record}
\label{structure}
\end{figure}

Figure \ref{structure} shows how the patient's data and metadata are organized: the patients' data are stored off-chain: locally (in the clinician database) and in the cloud as presented in Figure \ref{structure} (a) based on their categories. Currently we use three categories in our prototype: History and physical exams, Laboratory results, and Delivered radiation doses. In the future, we plan to define categories based on both the semantics and the sensitivity level of the data. Data files related to the patient and uploaded by different clinicians are stored within the corresponding category. A patient could optionally store some private data or notes encrypted with the patient public key, $\pks_{P}$.

Figure \ref{structure} (b) presents the structure of the patient's metadata
that consists of the following blocks: Permissions, Clinical Metadata, and Patient Private Data (optional). 
Permissions block is organized as follows: every Permission corresponds to a Doctor Id, with which a clinician is registered in the system. 
Every permission specifies the timeframe (\texttt{From: To:}) during which a clinician has a \texttt{Right} to \textbf{read} the patient's data that fall into a specific \texttt{Data Category}, upload them to the cloud repository (\textbf{write}), or \textbf{share} the patient's data within a framework of a specific research study, \texttt{Study Id}. For the latter the patient could also use \texttt{Anonymity} tag to specify if the data must be anonymized before sharing or could be shared as they are. \texttt{Timestamp} makes every permission unique and allows a patient to update and track access control changes corresponding to the same Doctor Id.


Clinical Metadata are a block that contains information about all the data files uploaded to the cloud by different clinicians. The Metadata Items are categorized based on the semantics of the corresponding data files. Every Item contains an Id of the clinician that uploaded the data (\texttt{Doctor Id}), a pointer to the file that is stored in the cloud, \texttt{Path to File}, the Hash of the Data File, \texttt{Hash(File)}, to ensure unforgeability of the data stored in the cloud, and the \texttt{Timestamp} of the event when the Data File was uploaded.
Similarly to the Patient Private Data stored on the cloud, some private data could be added by a patient to be stored in the State associated with the chaincode (CC).
The metadata are stored as a ``value'' part of a CC State. It can be accessed and modified using the functions that can be invoked on a CC.

To ensure a correct functioning of the developed chaincode we built a network that consists of a Membership service and four Nodes capable of running PBFT consensus protocol. Four is the minimum number of nodes needed to run the PBFT consensus protocol. We deployed CC on every node and issued a set of the ``invoke'' transactions 
(that trigger creation of a new patient metadata record, adding a permission, and uploading the metadata item), and ``query'' transactions to access the information from the State.
Currently, a patient is able to create a metadata record on the chaincode, add permissions, and retrieve his up-to-date metadata record, and, thus, his data that are stored on the cloud. A user registered with a Doctor role is able to upload, access and share for research purposes the data in the cloud based on the permissions specified by the patient.

Verification of the access control rights (currently \textbf{read}, \textbf{write} or \textbf{share}) is done via Logic of a chaincode written in Go programming language \cite{go}. For instance, every time a clinician would try to add new data on the cloud repository, a permission corresponding to this clinician has to be retrieved from the patient's metadata record. Then, the validity of the permission with respect to the data category and timeframe is controlled. Similarly, sharing patient data for the research purposes can not be performed by clinician without patient's agreement. This is guaranteed by the chaincode implementation.
Interfacing our system with the existing clinical database management systems and conducting more experiments with the data of the real patients are next steps of our work.

\section{Discussion}
Next we discuss the privacy, security, and scalability of the proposed framework.

\textit{\textbf{Privacy.}}  A patient's privacy is ensured by providing the patient with a possibility to specify  fine-grained access control over his data via permissions. 
Permissions are enforced by chaincode logic and, therefore, can not be violated by any user, unless the consensus protocols fails. The latter could happen only if a fraction of the verifying nodes intentionally tries to damage network operations. Centralized membership service already protects against Sybil attacks. Moreover, in the permissioned network, the nodes identities are known, therefore, there is no incentive for malicious behavior. 
In the case if a node still behaves maliciously, access to the network could be promptly restricted for this node.

Membership service also controls the identity of the users. Before registering a clinician, his identity is verified in the National Practitioner Data Bank.  A patient is registered with his $UII$, but all his data are linked to the pseudonym generated using his secret key, $\skaes_{P}$. Therefore, Membership service does not have an access to the patient's clinical data, yet guarantees authenticity of the users (via digital signature verification). If $\skaes_{P}$ is compromised or lost, access to the network will be recovered using $UII$ of a patient, a new key will be generated, and proxy reencryption \cite{Ateniese:2006} will be used.


\textit{\textbf{Security.}}
Clinical data stored in the cloud repository are encrypted with a patient secret key, $\skaes_{P}$ to provide data \textit{confidentiality}. Only the patient can share encryption key and set up the access control policy via permissions.
Shared data from the cloud registry are hashed and signed with a secret key of a user ($\sks_{U}$), before the data are uploaded. The hashes are stored as a part of a corresponding metadata item in the State. Transactions are also digitally signed, thus the data \textit{integrity} is ensured.

\textit{Availability} of the shared data is guaranteed by providing a cloud platform to store the data. Role-based APIs can be used at any node registered in the network to invoke or query the chaincode. As already mentioned, if a patient loses his credentials, access to the data stored on- and off-chain could still be recovered.

\textit{\textbf{Scalability.}} Clinical data sharing requires scalability of the system in terms of both the number of users and the number of nodes. PBFT consensus protocol provides excellent scalability in terms of the number of users, but have not  been well explored in terms of the number of Nodes (verified only up to few tens of Nodes) \cite{vukolic2015quest}. Possible scalability issues could be addressed by using hierarchical BFT protocols. Frequency of creating a block or number of transactions in a block (batch size) could be adjusted. System load is already minimized by storing patient's clinical records off-chain. We plan to evaluate the system performance and scalability in clinical settings in future work.

\section{Related Work}
The potential of the applications built on top of the blockchain technology for healthcare data management has been recently discussed \cite{MedRec,yue2016healthcare,jenkins2015bio,beninger2016pharmacovigilance}. Yue et al. claim to be the first to import blockchain into the design of a healthcare system \cite{yue2016healthcare}. They presented the architecture of a healthcare data gateway application for easy and secure control and sharing of medical data between different entities that may use patient data. However, the system has not been implemented nor tested yet. A possibility of sharing the data for research purposes is only sketched in the paper, without any security or privacy evaluation. Jenkins et al. proposed to use blockchain technology for a multifactor authentication in a specific research scenario (medical large data analysis with functional biomarkers) that involves biometric and biomedical data \cite{jenkins2015bio}.

MedRec \cite{MedRec} is the first and the only functioning prototype that have been proposed until now. The authors presented a system based on Ethereum smart contracts for an intelligent representations of existing medical records that are stored within individual nodes on the network. Two incentivizing models for ``mining'' are also proposed in \cite{MedRec}. Our prototype significantly differs from the framework in \cite{MedRec}. First,
MedRec is based on permissionless blockchain implementation and PoW, thus involves transaction fees, and requires involvement into ``mining'' and account management processes. In contrast, we have chosen permissioned blockchain implementation based on the requirements from the medical perspective. We justified our choice in Section 3.4. Second, in \cite{MedRec} the patients data are stored locally at every node. We decided to use a cloud-based storage, and employ encryption and key-sharing to ensure availability of the data even if the hospital node is temporary off-line.

\section*{Conclusion and Future Work}
In this paper, we proposed scenarios of blockchain technology application in different healthcare settings: primary care, medical data research, and connected health. We discussed how maintaining an immutable and transparent ledger, which keeps track of all the events happened across the network, could improve and facilitate the management of medical data.

Based on the constrains related to the healthcare context, we justified the choice of the permissioned blockchain technology for the implementation of the proposed scenarios.
We also presented an architecture of the framework for the specific needs in case of radiation oncology data sharing and implemented a prototype that ensures privacy, security, availability, and fine-grained access control over highly sensitive patients' data.

As part of future work, we would like to extend the structure of a patient record and its metadata, using the semantics of healthcare data, including the possibility of sharing  radiology images, which is much more challenging.  Since we work in collaboration with a hospital, we plan to test our system with the data of the real patients.
Our long term goal is to explore other scenarios proposed in the paper (such as connected health and medical data research) 
and apply them in practice to enhance the current healthcare data management.

\makeatletter
\renewcommand{\@biblabel}[1]{\hfill #1.}
\makeatother

\bibliographystyle{unsrt}

%




\end{document}